\documentclass[amssymb,useAMS,prd,aps,amsmath,nofootinbib,superscriptaddress,
twocolumn]{revtex4-1}
\usepackage{natbib}
\usepackage{color}
\usepackage{graphicx}
\usepackage{amsmath}
\usepackage[caption=false]{subfig}
\usepackage[colorlinks=true,linkcolor=blue,citecolor=blue,urlcolor=black]{
hyperref}

\begin{document}
\title{How well can new particles interacting with neutrinos be constrained after a galactic supernova?}
\author{Jonathan H. Davis}
\affiliation{Theoretical Particle Physics and Cosmology, Department of Physics, King's College London, London WC2R 2LS, United Kingdom
\\ {\smallskip \tt  \href{mailto:jonathan.davis@kcl.ac.uk}{jonathan.davis@kcl.ac.uk}}}
\date{\today}
\begin{abstract}
A supernova event in our own galaxy will result in a large number of neutrinos detected on Earth within the time-frame of a few seconds. These neutrinos will have been produced thermally 
with, in principle, three distinct temperatures for the electron, anti-electron and remaining heavy flavours respectively. We revisit the possibility that new MeV-mass particles $\chi$ are also
produced thermally during the event, which scatter with the neutrinos and alter their temperatures. Our main emphasis is on the detectability of this effect using the neutrino spectrum,
given the large uncertainty on the temperature and density profiles of the stellar matter.
By marginalising over the parameters of a simple analytic model for the stellar profile, we find that Super Kamiokande could place an upper limit on the scattering cross section at the level of $\sigma_{\chi \nu} \sim 10^{-40} \cdot (T / \mathrm{MeV})^2$~cm$^2$ for 10~MeV mass $\chi$ particles at $90\%$ confidence.
A direct-detection-like experiment would be less susceptible to systematic uncertainties in the neutrino production and mixing, but this would need a target mass around 
100 tonnes in order to acquire enough statistics to compete with Super Kamiokande.
\end{abstract}
\maketitle

\section{Introduction}
A supernova (SN) event in our own galaxy will lead to a large flux of neutrinos arriving at the Earth within the space of a few seconds. The only such event detected so far has been Supernova 1987a~\cite{PhysRevLett.58.1490,Yuksel:2007mn,Burrows:1992ec}, at a time when neutrino detectors were considerably
smaller and less advanced than they are today. Hence the detection of another such event, which should occur within the near future~\cite{1970A&A.....7...59C}, will result in a great deal of new information regarding both supernovae and neutrinos~\cite{Sigl:1994da,Keil:2002in,Yuksel:2007mn,Chiu:2013dya,PhysRevD.54.1194,Burrows:1992ec,Dasgupta:2011wg}.

The majority of these neutrinos are produced thermally during the supernova event, where temperatures can reach around 10~MeV~\cite{Sigl:1994da,Keil:2002in,Yuksel:2007mn}, and carry away most of the released energy. They are emitted with a distribution with is approximately of the Fermi-Dirac form, with a temperature which differs for the neutrino flavours. Broadly the electron neutrinos
should have the lowest temperature as they scatter the most in the star and so decouple at the largest radius. By contrast electron anti-neutrinos scatter less with the stellar matter and so
are emitted at a higher temperature, and the remaining heavy flavours have a higher temperature still. 

Various authors have demonstrated that, in addition to neutrinos, other more exotic particles could be produced thermally during the supernova event~\cite{Sigl:1994da,Fayet:2006sa,Goudelis:2015wpa}.
These particles $\chi$ are `dark' in the sense that they interact only weakly with visible matter and so are difficult to detect, except perhaps in extreme environments such as a hot post-supernova star.
Indeed if such particles interact with neutrinos through scattering or annihilation they will alter the various temperatures of the neutrino spectrum~\cite{Fayet:2006sa}, which could result in observable effects on Earth. 
However the exact temperatures are not well known and depend upon assumptions related to the structure of the 
 star during thermal production, specifically the dependence of density and temperature with radius~\cite{Sigl:1994da,Keil:2002in,Yuksel:2007mn,Raffelt:2001kv}.

\begin{figure}[t]
\centering
\includegraphics[width=0.49\textwidth]{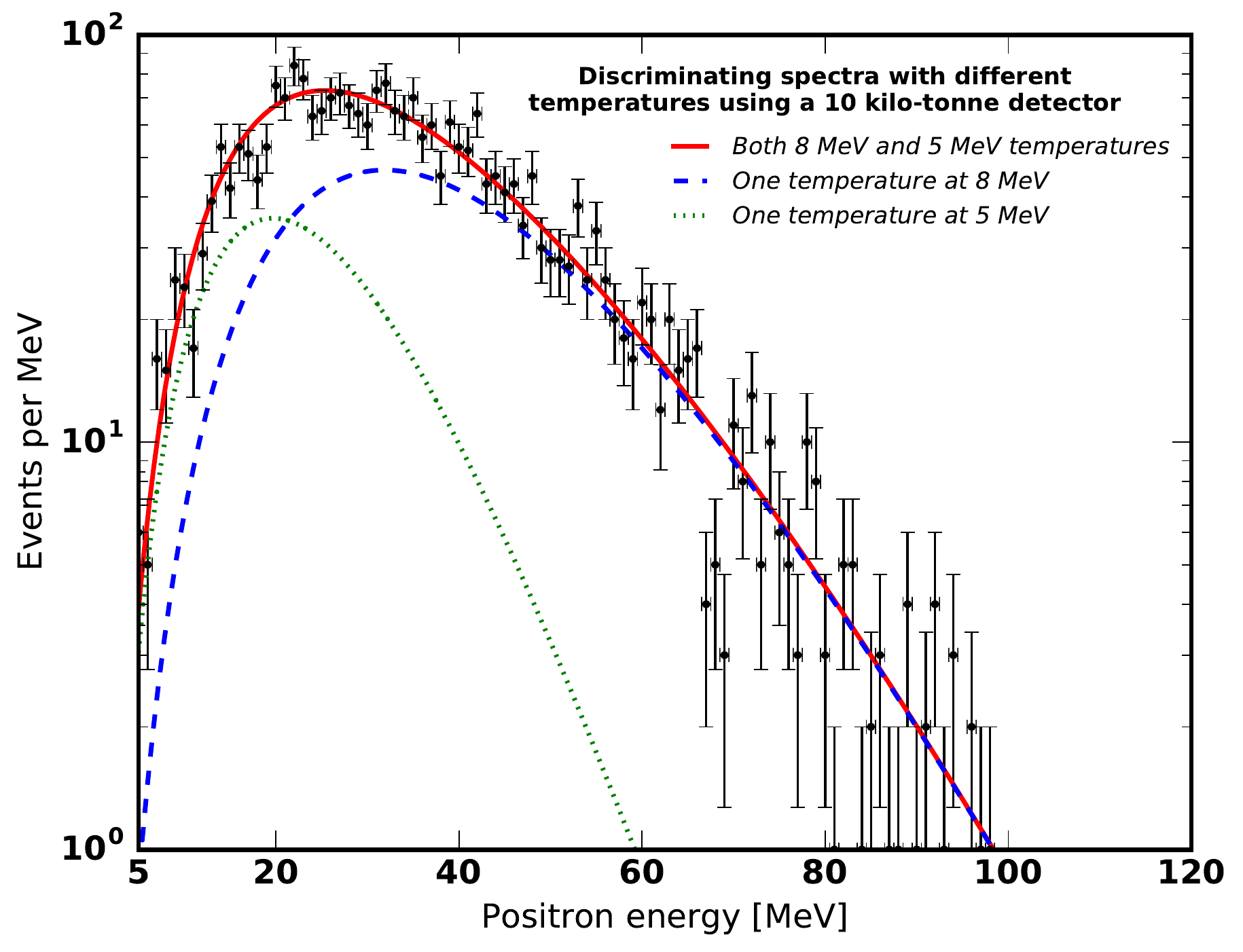}
\caption{Simulated data for a 10 kilo-tonne neutrino experiment which detects $\bar \nu_e$ from a supernova, whose spectrum is a combination of Fermi-Dirac distributions with two temperatures at 5 MeV and 8 MeV. We compare this with
spectra with only a single temperature of either 5 MeV and 8 MeV and the combined spectrum.}
\label{fig:spectra}
\end{figure}

In this work we seek to understand how the scattering with neutrinos of additional thermally-produced MeV-mass particles  will alter the neutrino temperatures, given the potentially large uncertainties in the modelling
of the star during the supernova event. 
Using this information we quantify to what extent current and future experiments can exploit measurements of the neutrino temperatures to  constrain the existence of such particles. 
We focus specifically on the difference between the temperatures of the neutrino flavours,
since dark sector scattering with neutrinos drives the temperatures of the different flavours towards a single value. This is important since, as shown in figure~\ref{fig:spectra}, a spectrum which is the sum of two (or more) different temperatures is clearly distinct from one which is made up of only
one temperature. 

In section~\ref{sec:production} we outline how the  temperatures of the neutrino species are calculated and in section~\ref{sec:chi} we describe how these are modified in the presence of new states $\chi$. In section~\ref{sec:detection} we give the formulae for the 
neutrino flux spectrum at Earth from a galactic supernova and how this is converted to the spectrum in a detector. We use these formulae in section~\ref{sec:cons} to set projected upper limits on the scattering between $\chi$ and 
neutrinos using a Bayesian analysis and compare these to other constraints in section~\ref{sec:comparison}. We conclude in section~\ref{sec:conc}.

\section{Thermal production of neutrinos and the neutrino temperature \label{sec:production}}
In this section we outline how the different neutrino temperatures $T$ and number densities $n$ are calculated using the diffusion equation, followed in section~\ref{sec:chi} where we describe how these are affected by scattering with a new `dark' particle $\chi$. 
Our approach is approximate, and we do not claim accuracy on par with numerical simulations of the supernova event~\cite{PhysRevLett.104.251101,Keil:2002in,Buras:2005rp,Mirizzi:2015eza,Raffelt:2001kv},
however it is sufficient for our purpose of understanding how sensitive the different neutrino temperatures are to modelling uncertainties in the star during thermal production. 
The diffusion equation takes the form~\cite{Sigl:1994da},
\begin{equation}
D n^{\prime \prime} + \left(D^{\prime} + \frac{2}{r} D\right) n^{\prime} = 
\sigma v (n^2 - n_{\mathrm{eq}}^2) + \frac{n - n_{\mathrm{eq}}}{\gamma \tau} ,
\label{eqn:diff}
\end{equation}
where $r$ is the radial distance from the centre of the star, a prime denotes an $r$-derivative, $D$ is the diffusion coefficient, $\sigma v$ is the self-annihilation 
cross section for the particle, $n_{\mathrm{eq}}$ is the number density of the 
particle when it is in thermal equilibrium with the stellar matter, $\gamma$ is the Lorentz factor and $\tau$ is the lifetime of the particle.

The value of $r$ where $n$ departs from its equilibrium value i.e. where $n - 
n_{\mathrm{eq}} \sim n_{\mathrm{eq}}$ is called the neutrinosphere radius (in 
the case of neutrinos at least). This is effectively the radius where the neutrinos can be said to have decoupled from the stellar matter. Likewise the corresponding temperature $T$ of the 
star at this radius is the neutrinosphere temperature. The neutrinosphere
temperature determines the observed neutrino temperature on Earth from the supernova. This can be 
approximated by reducing equation~\eqref{eqn:diff} to the following expression~\cite{Sigl:1994da},
\begin{equation}
\left|D n^{\prime \prime}_{\mathrm{eq}} + \left(D^{\prime} + \frac{2}{r} D\right) n^{\prime}_{\mathrm{eq}} \right| =
\left|\sigma v n_{\mathrm{eq}}^2 + \frac{n_{\mathrm{eq}}}{\gamma \tau} \right| ,
\label{eqn:thermal_decoupling}
\end{equation}
and solving for $T$, where we assume that $n_{\mathrm{eq}}$ is,
\begin{equation}
n_{\mathrm{eq}} \approx \left[\frac{3 \zeta(3)}{4 \pi^2} T^3(r) + \left(\frac{m T(r)}{2 \pi} \right)^{3/2} \right] \mathrm{exp} \left(\frac{- m}{T(r)} \right),
\label{eqn:equib}
\end{equation}
where $m$ is the mass of the particle (e.g. a neutrino or dark particle $\chi$).

We approximate the radial dependence of the baryon density and temperature of the star after the supernova event
using a simple empirical model, whose parameters we can vary easily to understand the effect of their uncertainty on the different neutrino flavour temperatures. As such we assume the following power-law forms
for the baryon density $n_B$ and the baryon temperature $T$,
\begin{eqnarray}
n_B(r) &=& n_0 \left( \frac{r_0}{r} \right)^k \label{eqn:nbprof} \\
T(r) &=& T_0 \left( \frac{r_0}{r} \right)^q \label{eqn:Tprof}
\end{eqnarray}
where $n_0$, $r_0$, $T_0$, $q$ and $k$ are in principle all free parameters. Some particular models take for example: $k = 5$, $q = 5/3$, $T_0 = 5$~MeV, $r_0 = 30$~km and 
$n_0 = 5.97 \cdot 10^{35}$~cm$^{-3}$~\cite{Sigl:1994da} (Model A) or $k = 10$, $q = 2.5$, $T_0 = 31.66$~MeV, $r_0 = 10$~km and 
$n_0 = 1.21 \cdot 10^{38}$~cm$^{-3}$~\cite{Keil:2002in} (Model B). For our projected limits we will treat these variables as nuisance parameters, since the exact form of the profile is not known and can vary significantly.

The value of the neutrinosphere temperature is fixed by the scattering cross section between
neutrinos and the other particles of the star, particularly baryons, and the self-annihilation
cross section of the neutrinos. The former effect arises through the diffusion coefficient,
which if we assume dominant scattering from baryons equals $D = v / (3 \sum_i f_{np,i} n_B \sigma_{B,i})$~\cite{Sigl:1994da,Fayet:2006sa}, where $v$ is the velocity of the neutrinos,
$\sigma_{B,i}$ is the scattering cross section with baryons through the channel labelled by $i$ (e.g. neutral-current) and $f_{np,i}$ is a factor relating to the relative abundance of neutrons and protons. 

Electron neutrinos scatter the most frequently and so have the smallest value of $D$ due to their interaction with neutrons via the charged current interaction as well as the sub-dominant neutral current channel. Electron anti-neutrinos 
by contrast scatter with protons via charged current interactions instead, whose density is several orders of magnitude smaller than neutrons in the star, resulting in a larger value of $D$. The remaining `heavy' flavours
have the largest $D$ and so the largest neutrinosphere temperature as they dominantly scatter with neutrons and protons via neutral current interactions, whose cross section is smaller than that for charged-current scattering. 
This all holds to first order, however effects from more detailed modelling of the supernova can alter this simple picture~\cite{PhysRevLett.104.251101,Keil:2002in,Buras:2005rp,Mirizzi:2015eza,Raffelt:2001kv}.

\section{Modification of the neutrino temperatures from scattering with $\chi$ \label{sec:chi}}

We now consider the possibility that an additional species of particle $\chi$, which interacts only weakly with the visible sector, is produced thermally in the supernova and scatters with neutrinos, thereby contributing to the diffusion coefficient $D$. 
Under the assumption that such new
interactions are flavour-blind the diffusion coefficient becomes,
\begin{equation}
D = v \cdot \left[3 \left(\sum_i f_{np,i} n_B \sigma_{B,i} + n_{\chi} \sigma_{\chi \nu}  \right) \right]^{-1} ,
\label{eqn:diff_chi}
\end{equation}
where $n_{\chi}$ is the number density of the new species $\chi$ and $\sigma_{\chi \nu}$ is its scattering cross section with neutrinos. Importantly if $n_{\chi} \sigma_{\chi \nu} \gg \sum_i f_{np,i} n_B \sigma_{B,i}$ then the
value of $D$ will be dominated by the flavour-blind scattering with the dark sector $\chi$, and so we might expect the neutrino temperatures to approach a single value regardless of flavour. 

The exact temperature dependence of $\sigma_{\chi \nu}$ depends on how the $\chi$ particles interact with neutrinos. For example following refs.~\cite{Mangano:2006mp,Boehm:2006mi} if one assumes either scalar $\chi$ interacting with neutrinos
via the exchange of a new fermion, or fermionic $\chi$ interacting with neutrinos via a new gauge boson, then $\sigma_{\chi \nu} \sim T^2$ is a reasonable assumption provided that the masses of $\chi$ and the exchange particle
are not degenerate. 
For the former case of scalar $\chi$ we envisage a term in the Lagrangian of form $$\mathcal{L}_{\mathrm{int}} = g_{\chi} F \chi \nu_L $$ where $F$ is a heavy right-handed fermion which mediates the interaction between $\chi$
and neutrinos. This then leads to a thermally averaged scattering cross section of form $\sigma_{\chi \nu} v \propto (g_{\chi}^4 / (m_{\chi}^2 - m_F^2)^2) \cdot T^2$. Such an interaction term could also result in the generation of a mass term for the neutrinos in analogue to the see-saw mechanism (see e.g. ref.~\cite{Boehm:2006mi}).
Hence our projected constraints are partly motivated as a method of constraining these neutrino mass models, however they can be interpreted in the context of other models and  we have checked our results in the case of alternative assumptions, such as $\sigma_{\chi \nu} \sim T$, and find no significant deviation from the conclusions presented here. 

Furthermore the form of $n_{\chi}$ depends on the interactions of the $\chi$ particle with nucleons and electrons as well as neutrinos. In this work we assume that the $\chi$
are in thermal equilibrium up to a radius larger than the neutrinosphere for all flavours. This means that $\chi$ has either a large annihilation cross section or decay rate and/or interactions with baryons in order to keep the $\chi$
coupled to the hot stellar matter. We return to this assumption in section~\ref{sec:comparison}. We restrict our analysis to particles with mass $m_{\chi} \gtrsim 10$~MeV to avoid constraints from nucleosynthesis
on thermally produced light particles~\cite{PhysRevD.70.043526}. This also means that the density of $\chi$ will be Boltzmann suppressed compared to that for neutrinos, meaning that the latter still dominate the energy transport
from the supernova (see section~\ref{sec:comparison} for further discussion).

\begin{figure}[t]
\centering
\includegraphics[width=0.49\textwidth]{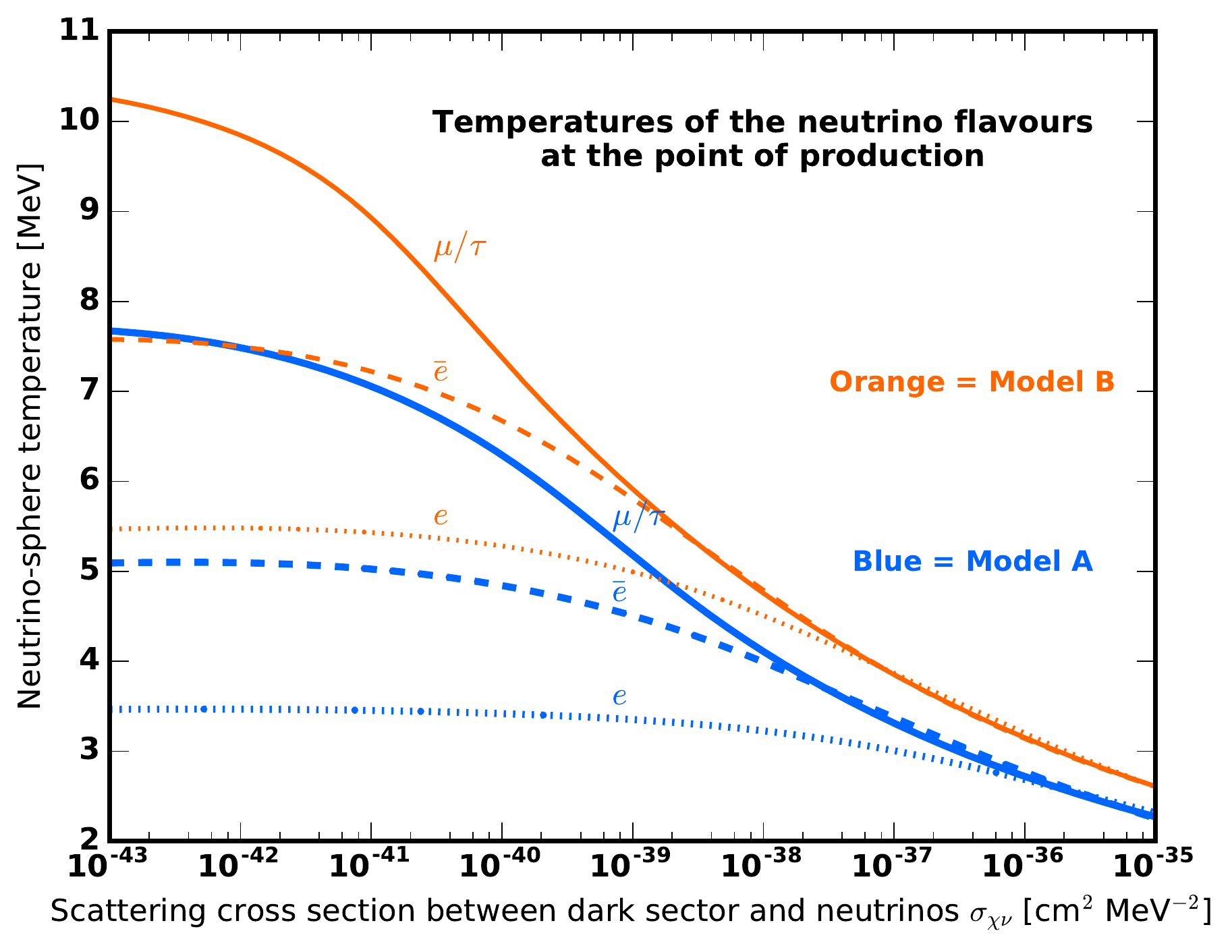}
\caption{Dependence of the temperature of the three different neutrino flavours at the point of production on the scattering cross section with dark sector particles $\chi$ (of form $\sigma_{\chi \nu} T^2$) with a mass of 10~MeV. 
The dotted line represents $\nu_e$, the dashed represents $\bar \nu_e$ and the solid line is for the remaining heavy flavours.
We show such dependence for two different assumptions on the profile
of the matter in the star during the supernova event, labelled as Model A and Model B.}
\label{fig:temps_plot}
\end{figure}

We show in figure~\ref{fig:temps_plot} the values of the three different neutrino temperatures as a function of $\sigma_{\chi \nu}$ for the two different example empirical models of the star introduced earlier as Model A and Model B.
These have been calculated using equation~(\ref{eqn:thermal_decoupling}) with $D$ from equation~(\ref{eqn:diff_chi}).
There are two main points to note here: the first
is that the values of the neutrino temperatures depend strongly on the choice of empirical model for the star during the supernova event. The second is that when $\sigma_{\chi \nu}$ is large enough the values of the temperatures
for each flavour become equal, and that this value of $\sigma_{\chi \nu}$ is more robust to changes in the empirical model of $n_B(r)$ and $T(r)$ than the absolute temperatures themselves.
Therefore if we are to obtain anything useful from analysing the temperatures of neutrinos from a supernova it will likely come from comparing the relative temperatures between species, rather than their absolute values. 
\emph{Hence although the values of the temperatures depend strongly on the model of the hot star for a given value of $\sigma_{\chi \nu}$, the number of distinct temperatures which make up the total neutrino spectrum is much less sensitive.}
In section~\ref{sec:cons}
we use this fact to understand what can be learned in future detectors by determining the number of different temperatures in the overall spectrum.

\section{Measuring the spectrum of supernova neutrinos on Earth \label{sec:detection}}
In this section we outline how neutrinos from a supernova will be detected on Earth using either inverse-beta decay capture of electron anti-neutrinos~\cite{Yuksel:2007mn} in experiments such as Super-Kamiokande~\cite{Fukuda2003418} or Hyper-Kamiokande~\cite{Abe:2011ts} or via
coherent nuclear recoils in direct detection experiments~\cite{Baudis:2012bc,Baudis:2013qla,Billard:2013qya,Monroe:2007xp}. The former has the advantage of greater statistics, while the latter is insensitive to the mixing between neutrino flavours on their way to Earth and allows for all three temperatures to in principle be measured.

The expected spectrum of neutrinos $\phi(E)$ from a supernova takes the form of a pinched Fermi-Dirac distribution~\cite{Vaananen:2011bf} (neglecting features from mixing effects~\cite{Dasgupta:2008cd,Dasgupta:2009mg}),
\begin{equation}
\phi(E_{\nu},\eta) = \sum_{i=1}^{N_{\mathrm{sp}}} \frac{R_i}{T_i^3 F_2(\eta)}  \frac{E_{\nu}^2}{\mathrm{exp}(E_{\nu}/T_i - \eta) + 1}  ,
\end{equation}
where $T_i$ is the temperature of the particular neutrino species (of which there are a number $N_{\mathrm{sp}}$), $R_i$ is the relative luminosity of species $i$ compared to the total, $\eta$ is the pinching parameter, which we assume to be zero in this work for simplicity, and $F_n$ is a Dirac integral, taking the form
\begin{equation}
F_n(\eta) = \int_0^{\infty} \frac{x^n \mathrm{d}x}{\mathrm{exp}(x-\eta) + 1} .
\end{equation}
Multiplying $\phi(E,\eta)$ by the total flux from a galactic supernova $\Phi_0$ gives the flux spectrum on Earth. We estimate the total flux using the time-integrated supernova luminosity $L_0 \sim 10^{53}$~ergs~\cite{Yuksel:2007mn}, the average
energy of each flavour $\langle E_i \rangle$ and the distance to the supernova
$d_0 \sim 10$~kpc (i.e. close to the galactic centre). Hence the total integrated flux spectrum of neutrinos from a galactic supernova takes the form,
\begin{equation}
\Phi(E_{\nu},\eta) = \sum_{i=1}^{N_{\mathrm{sp}}} \frac{L_0}{\langle E_i \rangle 4 \pi d_0^2} \frac{R_i}{T_i^3 F_2(\eta)}  \frac{E_{\nu}^2}{\mathrm{exp}(E_{\nu}/T_i - \eta) + 1}  .
\end{equation}

Detectors such as Super-Kamiokande and Hyper-Kamiokande~\cite{Fukuda2003418,Abe:2011ts} will detect supernova neutrinos primarily via inverse-beta decay (i.e. $\bar \nu_e + p \rightarrow n + e^+$) due to the large cross section for this process~\cite{Yuksel:2007mn,Dasgupta:2011wg}.
The events are identified by positrons whose energies $E_r$ are related simply to the neutrino energies $E_{\nu}$ as $E_r = E_{\nu} - 1.29 \mathrm{MeV}$.
The differential detection rate is then the product of $\Phi(E_{\nu},\eta)$ with the cross section for inverse beta decay. However the calculation is complicated by the fact that we only detect $\bar \nu_e$ which will be made
up of neutrinos which were in various different flavours at the point of production. Hence the spectrum of the $\bar \nu_e$ at Earth should be composed of two distinct temperatures provided that $\sigma_{\chi \nu}$ is negligible
and that the different anti-flavours mix on their way to Earth. This also means that the temperature of the $\nu_e$ flavour is not measurable with this detection method, and so there will only at most be two temperatures.
Hence when we determine projected limits on $\sigma_{\chi \nu}$ we need to take account of the potentially uncertain values of the relative luminosities $R_i$ for the two remaining temperatures.

By contrast, the detection of neutral current scattering allows all of the neutrino temperatures to be detected regardless how how the neutrinos mix on their way to Earth. Direct detection experiments looking for dark matter could be sensitive
to supernova neutrinos via this channel. In this case the neutrinos scatter coherently on nuclei with a differential cross section ${\mathrm{d}\sigma}/{\mathrm{d}E_r}$~\cite{Billard:2013qya,Monroe:2007xp}
leading to a differential scattering rate of nuclear recoil events taking the form,
\begin{equation}
\frac{\mathrm{d}R}{\mathrm{d}E_r} = \int_{E_{\nu,\mathrm{min}}}^{\infty} \mathrm{d} E_{\nu} \frac{\mathrm{d}\sigma}{\mathrm{d}E_r}(E_r,E_{\nu}) \Phi(E_{\nu},\eta),
\end{equation}
where $E_{\nu,\mathrm{min}} = \sqrt{m_N E_r/2}$ is the minimum neutrino energy able to induce a recoil of energy $E_r$ for a nucleus of mass $m_N$. In this case since the detection method is via neutral-current scattering all three temperatures can be detected
and the relative luminosities $R_i$ are set at the point of production (ignoring possible new physics effects during transit).

\begin{figure}[t]
\centering
\includegraphics[width=0.49\textwidth]{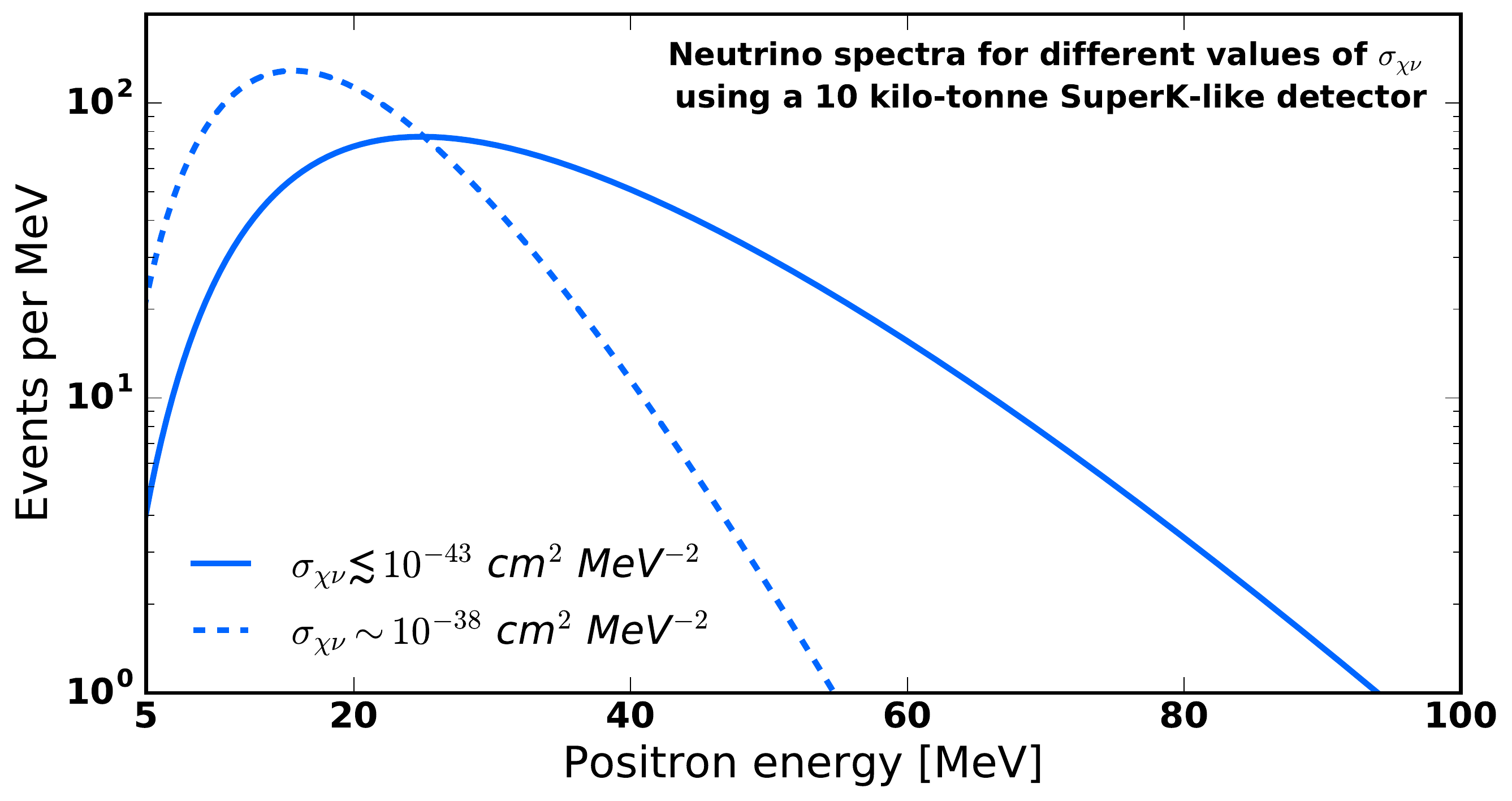} \\ \vspace{5pt}
\includegraphics[width=0.49\textwidth]{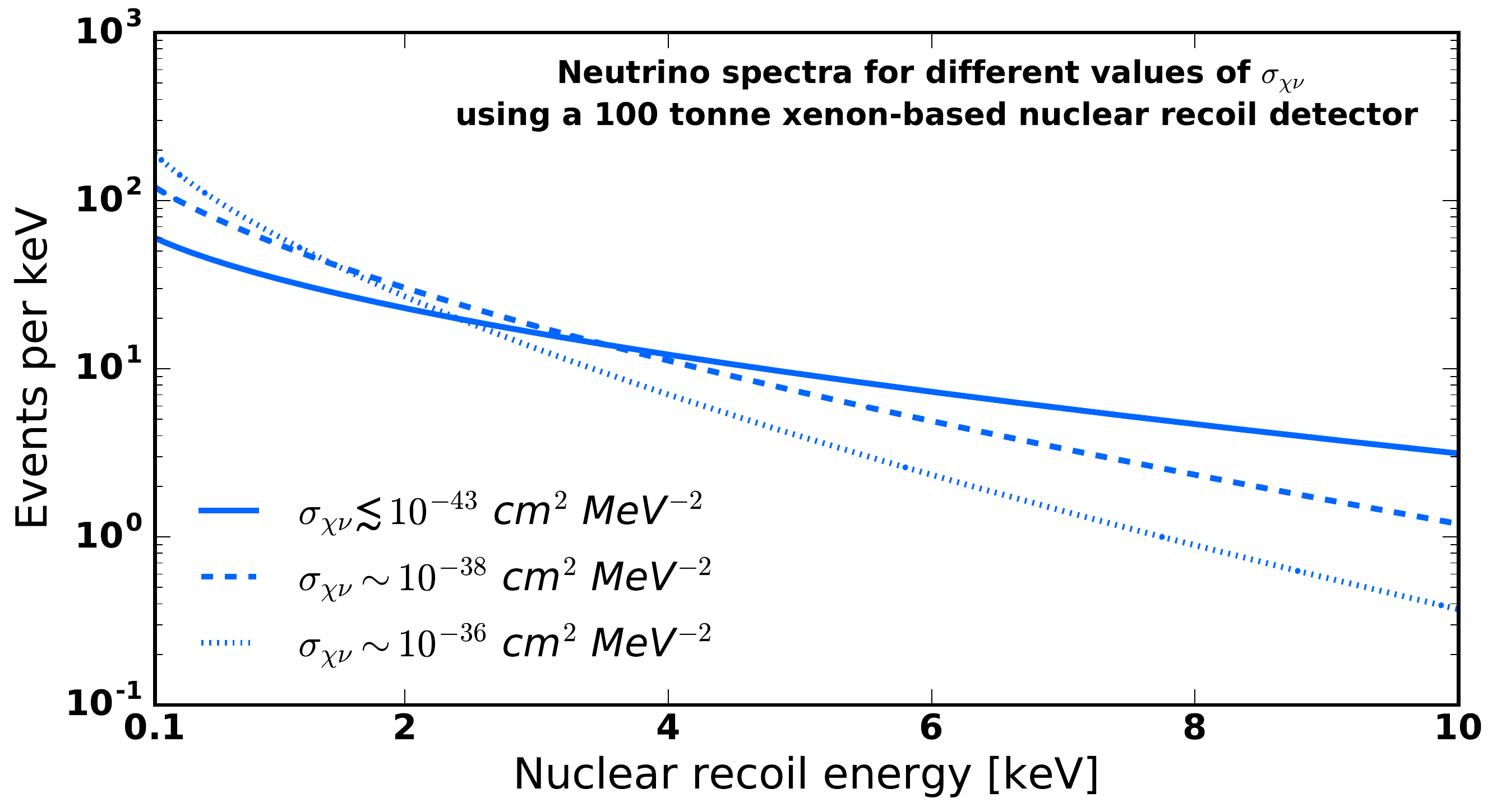}
\caption{Expected neutrino spectra for different values of the scattering cross section $\sigma_{\chi \nu}$ with dark sector particles $\chi$ in an experiment detecting neutrinos via inverse beta decay (similar to Super Kamiokande~\cite{Fukuda2003418}) or by nuclear scattering.
The different spectra correspond to the neutrino flavour temperatures predicted assuming Model A in figure~\ref{fig:temps_plot}.}
\label{fig:spectra_sigmas}
\end{figure}

Shown in figure~\ref{fig:spectra_sigmas} are the expected spectra
in both types of detector considered here, for different values of $\sigma_{\chi \nu}$. As $\sigma_{\chi \nu}$ increases the spectra shift towards having a single temperature as opposed to being composed of multiple distinct temperatures. 
This is most obvious with knowledge of the whole spectrum, and so
in order to discriminate between the different spectra we need to be able to observe both low and high energies events, and so a 
low threshold is crucial.

We are now in a position to derive projected exclusion limits on the dark sector interaction cross section with neutrinos $\sigma_{\chi \nu}$, using the spectra calculated in this section. How strong these limits can be will be determined by how effectively we can discriminate
the different constituent temperatures in the spectrum from the supernova, and the systematic uncertainties from the modelling of the star during the supernova event.

\section{Constraining new particle interactions with neutrinos after modelling uncertainties \label{sec:cons}}
\begin{figure*}[t]
\includegraphics[width=0.48\textwidth]{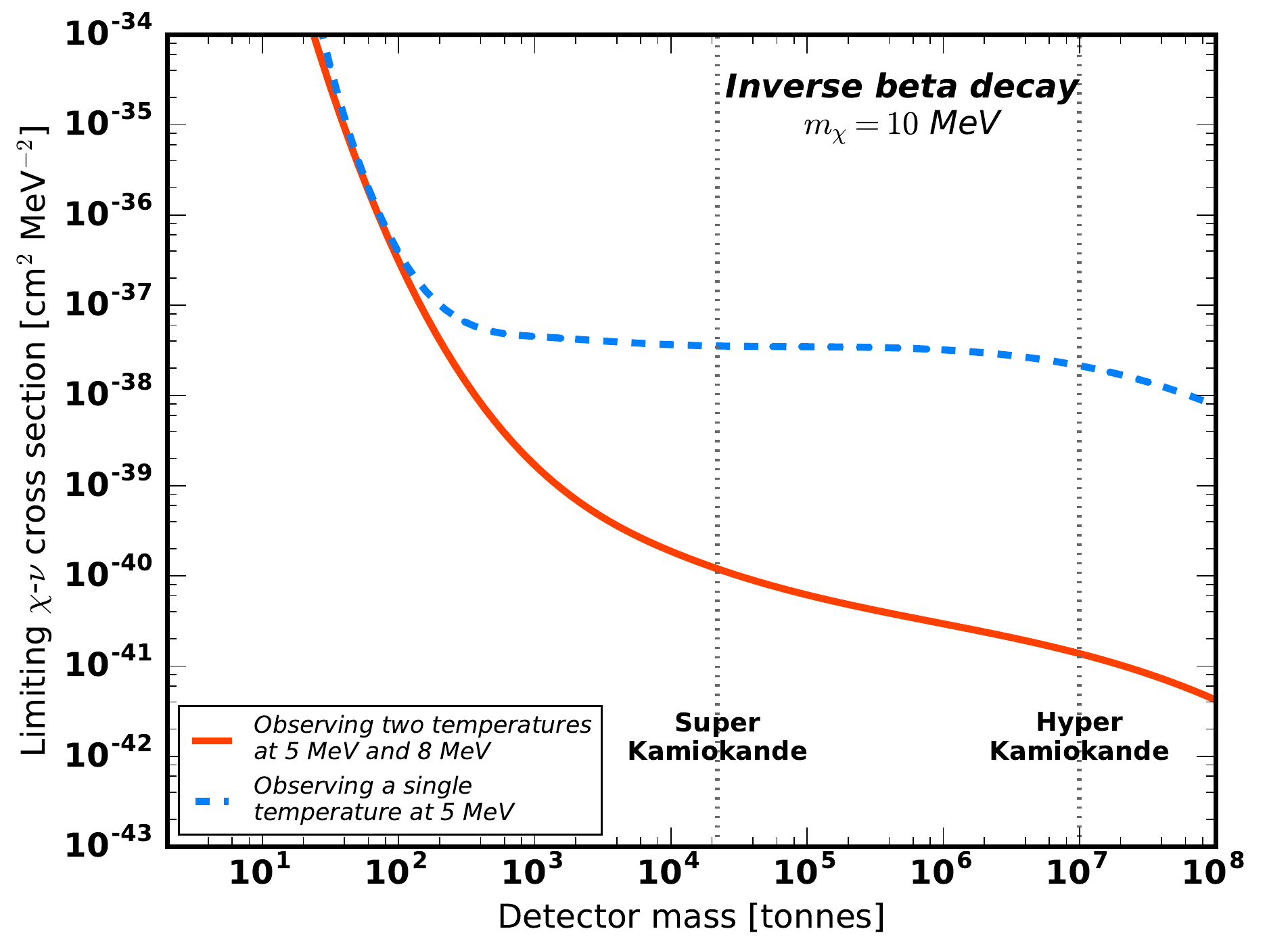} \hspace{0pt}
\includegraphics[width=0.48\textwidth]{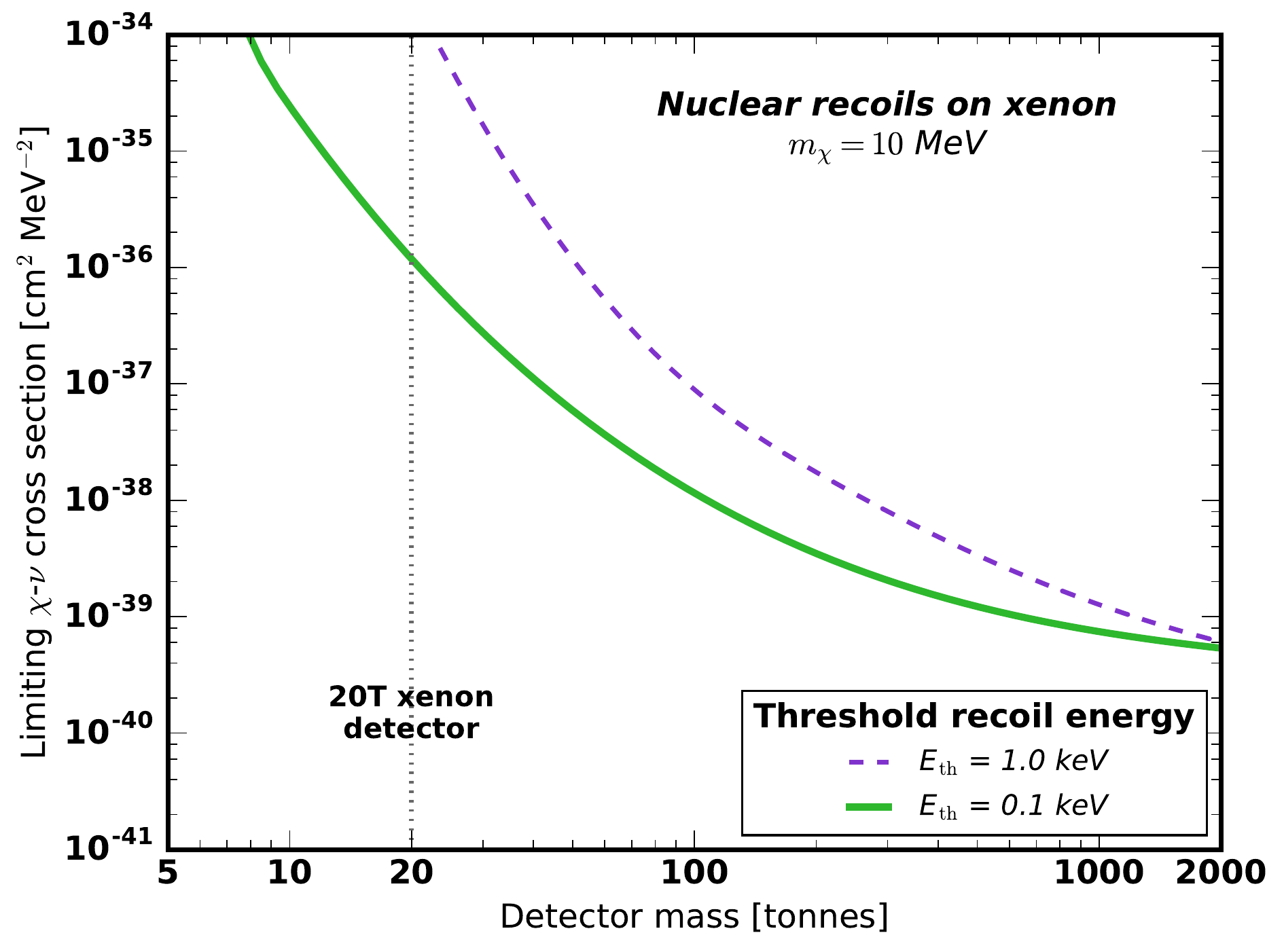}
\caption{Projected $90\%$ upper limits for the scattering between new particles $\chi$, produced thermally during a supernova, and neutrinos. We show such limits for various detector target masses and two distinct detection
mechanisms i.e. inverse beta decay and coherent nuclear scattering. For the former we consider limits in two different cases, depending on the spectrum observed from a future supernova.
These limits have been obtained after marginalising over the uncertainty on the neutrino temperatures from the modelling of the star during the supernova event.
The left plot assumes a low-energy threshold of 5~MeV for detection of positrons, while for the right plot we assume two different threshold energies $E_{\mathrm{th}}$ for nuclear recoils.}
\label{fig:lims}
\end{figure*}

In this section we determine the projected sensitivity of future multi-tonne experiments to the scattering cross section between dark sector particles and neutrinos, produced thermally in a galactic supernova. Such interactions
will alter the temperatures of the different neutrino flavours (see figure~\ref{fig:temps_plot}), however exactly how this happens depends crucially on how well we know the dependence of the baryon density and temperature of the star
during the supernova event. Hence our projections must incorporate this uncertainty in the choice of profile. 

To do this we perform a Bayesian analysis, while parametrising the empirical model of the hot star with nuisance parameters, which we marginalise over. For such parameters we use $k$, $q$ and $T_0$ from equations (\ref{eqn:nbprof}) and (\ref{eqn:Tprof}),
which determine how steeply the temperature and baryon density fall off with radius. The posterior density, used to determine the exclusion sensitivity for a particular data-set $d_j$, is
\begin{equation}
\begin{split}
\mathcal{P}(d_j,\sigma_{\chi \nu}) \propto \int \mathcal{L}(d_j,\sigma_{\chi \nu},k,q,T_0) \\
\mathcal{P}(k) \mathcal{P}(q)  \mathcal{P}(T_0)  \mathcal{P}(\sigma_{\chi \nu}) \, \mathrm{d} q \, \mathrm{d} k \, \mathrm{d} T_0 ,
\end{split}
\end{equation}
where $\mathcal{L}$ represents the likelihood function, which we assume to be of Poisson form to compare the expected spectrum for a given stellar profile and value of $\sigma_{\chi \nu}$ (e.g. figure~\ref{fig:spectra_sigmas}) to simulated data.
$\mathcal{P}$ are the prior distributions for the three nuisance parameters and the signal parameter $\sigma_{\chi \nu}$.
We then set a projected limit by generating a number of simulated data-sets $N_d$ given a particular detector mass and use the posterior summed over these data-sets
$\mathcal{P}(\sigma_{\chi \nu}) = \sum_j^{N_d} \mathcal{P}(d_j,\sigma_{\chi \nu})$. We generate these data-sets by assuming a particular set of temperatures and relative luminosities for each flavour.
Integrating $\mathcal{P}(\sigma_{\chi \nu})$ from $\sigma_{\chi \nu} = 0$ up to $90\%$ of the total posterior probability then gives us our projected exclusion limits.

The choice of priors for the nuisance parameters represents how well these quantities, and thus the profile of the baryonic matter during neutrino (and $\chi$) thermal production, are known. Since we are interested in understanding how well $\sigma_{\chi \nu}$
can be probed even if the stellar profile is poorly known we choose priors which cover a wide-range of $q$, $k$ and $T_0$. In principle however the profile uncertainty could be reduced using e.g. simulations of the supernova event~\cite{Keil:2002in,Buras:2005rp,Mirizzi:2015eza}.

Figure~\ref{fig:lims} shows projected limits for either detection via inverse-beta decay or coherent
nuclear scattering (other detection methods have also been proposed e.g. ref.~\cite{Vaananen:2011bf}) assuming priors which equal a constant in the following ranges (and zero otherwise),
\begin{eqnarray}
\mathcal{P}(k) &\in& [0,10] \\
\mathcal{P}(q) &\in& [1,10] \\
\mathcal{P}(T_0) &\in& [2,20] \, \mathrm{MeV},
\end{eqnarray}
with $r_0$ and $n_0$ fixed at values of $r_0 = 30$~km and $n_0 = 5.97 \cdot 10^{35}$~cm$^{-3}$. The choice of these priors is arbitrary, and we have intentionally chosen them to represent a large degree of uncertainty by spanning a wide range of possible stellar profiles, including
the values of $k$, $q$ and $T_0$ from model A and model B introduced earlier. Greater knowledge of the profile of the star during neutrino production would reduce the range of these priors and so would likely strengthen our projected limits.

In order to determine these projected limits we have simulated datasets in detectors of various target masses. Since we do not know \emph{a priori} what spectra they will observe, depending on the physics during the supernova and the
mixing of the neutrino flavours, we have set limits under two different assumptions. For example in the left panel of figure~\ref{fig:lims} we set limits 
under the assumption that the spectrum of $\bar \nu_e$ from the supernova event is either made up of a single Fermi-Dirac spectrum of temperature 5~MeV or of two spectra with temperatures of 5~MeV and 8~MeV. For the former case we do
not know whether $\chi$-neutrino scattering has forced the temperatures of the different anti-neutrino flavours together, whether a different mechanism has brought the temperatures together, or whether the electron anti-neutrino flux is dominated by neutrinos which were produced in that same
flavour at the supernova. Hence conservatively all one can do is set an upper limit on $\sigma_{\chi \nu}$ at the level where it would drive the temperature of the neutrinos to be smaller than 5~MeV, given uncertainties on the 
stellar profile.

However by comparison the projected limit on $\sigma_{\chi \nu}$ will be much stronger if a spectrum with a combination of temperatures is observed (see also figure~\ref{fig:spectra}), which is the more likely scenario if the
neutrino flavours mix on their way to Earth.
In this case as can be seen from the left panel of figure~\ref{fig:lims} the exclusion
limit will be at the level of $\sigma_{\chi \nu} \sim 10^{-40} \cdot (T / \mathrm{MeV})^2$~cm$^2$ for Super Kamiokande~\cite{Fukuda2003418}, assuming two temperatures at 5~MeV and 8~MeV with equal relative luminosities. This is partly due to it being easier to discriminate different spectra with information on multiple temperatures, given the large uncertainties in the profile
of the baryons after the supernova event. Essentially with enough statistics a spectrum with two temperatures can be easily distinguished from one with only one temperature, even if the uncertainties on the values of the temperatures themselves
are significant.
The limit which Hyper Kamiokande~\cite{Abe:2011ts} should be able to set should be at least an order of magnitude stronger due to the larger number of events it will be able to observe.

We show in the right panel of figure~\ref{fig:lims} the projected limits for a xenon-based experiment which observes supernova neutrinos through coherent nuclear scattering with a combined spectrum of three temperatures
at 8~MeV, 5~MeV and 3.5~MeV, similar to the result for Model A in figure~\ref{fig:temps_plot}.
In this case there is no uncertainty from the mixing of the different neutrino flavours since the detection is via the neutral-current, however the smaller target masses of experiments which exploit this effect, mainly dark matter direct detection experiments,
means that they are unlikely to build up enough statistics to compete with Super or Hyper Kamiokande. Nevertheless a xenon-based detector with a mass around 20 tonnes (e.g. Darwin~\cite{Baudis:2012bc,Baudis:2013qla}) should be able to set a limit on the dark sector scattering cross section with neutrinos
provided the threshold for nuclear recoils can be lowered to around 0.1~keV. This may be possible if such detectors look only at ionisation signals, for which the threshold is generally lower, since discrimination between nuclear
and electronic recoils will not be required for such a significant event burst.

The key advantage of the direct detection experiments is their ability to measure all three potential temperatures, while relying on inverse beta decay capture of $\bar \nu_e$ allows only the temperatures of the anti-flavours to be
measured. This is important as any effect within the supernova which could drive the temperatures together without new physics, for example that of ref.~\cite{PhysRevLett.104.251101}, or an effect which suppresses the mixing of the neutrino
flavours in transit to Earth
would weaken our limits, especially in the case of
 $\bar \nu_e$ detection. Hence it is possible that the limit from coherent nuclear scattering is more robust to such effects, beyond the simplistic model of the star during the supernova event which we have used here. Indeed this motivates the 
 construction of such detectors up to 100 tonnes or more in mass, as an effective way of detecting all flavours of supernova neutrinos with high precision.
 
\section{Comparison with other limits on new MeV-mass particles \label{sec:comparison}}
We now consider such $\chi$ particles within a wider context by comparing our projected limits with other constraints. These particles should be only weakly coupled at most to visible matter to avoid having been already detected,
however even so there are other methods of probing their interactions.

Indeed the temperature of the neutrinos is not the only method of using a supernova to constrain new particles, which interact only weakly with visible matter. For example the $\chi$ particle needs to have a mean free path for scattering which is shorter than the size of the hot
star during the supernova event, to avoid excessively rapid cooling~\cite{Boehm:2013jpa} i.e. a new particle produced thermally but which scatters less often with the stellar matter than neutrinos do could make the energy loss rate too rapid.
Hence the $\chi$ particle likely needs interactions in addition to those with neutrinos in order to evade this cooling bound, and so our assumption of thermal equilibrium for $\chi$ is reasonable.

To study thermal equilibrium in more detail we follow a similar argument using equation~(\ref{eqn:diff}). Broadly speaking if $D$ is small then $n$ is close to $n_{\mathrm{eq}}$ and so the particles are in thermal equilibrium (or close to it).
In this case a small value of $D$ means that the $\chi$ particles scatter frequently with the stellar matter.
Indeed assuming that $\chi$ maintains its equilibrium through interactions with the baryonic matter of the star we can
make use of the mean free path for scattering $L_{\chi B}$ to estimate the required scattering cross section. We need that $L_{\chi B} \ll L_{\mathrm{SN}}$, where $L_{\mathrm{SN}}$ is the linear size of the hot star. Assuming that $L_{\chi B} \sim (n_B \sigma_{\chi B})^{-1}$, where $\sigma_{\chi B}$ is the scattering cross section between baryonic matter and $\chi$ we have
that $n_B \sigma_{\chi B} \, L_{\mathrm{SN}} \gg 1$. The value of $n_B$ depends on the model for the hot star i.e. equations (\ref{eqn:nbprof}) and (\ref{eqn:Tprof}) and the temperature. If we assume that we need the $\chi$ to be
in thermal equilibrium at temperatures above 1~MeV then using parameters from Model A we have that $n_B \sim 5 \cdot 10^{33}$~cm$^{-3}$. Hence if $L_{\mathrm{SN}} \sim 10$~km we have the approximate requirement that
$\sigma_{\chi B} \gtrsim 10^{-38}$~cm$^2$ for thermal equilibrium of the $\chi$ particles down to temperatures of 1~MeV. However this bound could be several orders of magnitude higher for more extreme models of the hot star during
neutrino thermal production.

Equation~(\ref{eqn:diff}) also tells us that the self-annihilation cross section of $\chi$ should be non-zero for thermal equilibrium to be established, though this depends partly on whether $\chi$ can decay. Assuming that $\chi$ has a 
lifetime much longer than the time-period for thermal production then the larger the value of $D$ the larger the self-annihilation cross section $\sigma v$ needs to be for $n$ to be equal (or close) to $n_{\mathrm{eq}}$. Solving
equation~(\ref{eqn:diff}) numerically for the number density $n$ of the $\chi$ particles with $\sigma_{\chi B} = 10^{-38}$~cm$^2$ confirms that $\chi$ are in thermal equilibrium down to temperatures of around 1~MeV for a self-annihilation
cross section $\sigma v \sim 10^{-26}$~cm$^3$s$^{-1}$, which is the scale required for $\chi$ to be the dark matter of the Universe. Values of $\sigma v$ larger than this are also allowed for thermal equilibrium, however in this case
the relic density of $\chi$ will be too small for it to be dark matter.

Bounds also exist on the temperature of $\nu_{\mu}$ and $\nu_{\tau}$ produced in supernovae from the abundances of light elements~\cite{PhysRevLett.94.231101}. This constraint is complimentary to the one studied in this work and could
in principle be used to constrain $\sigma_{\chi \nu}$.
As mentioned previously thermally produced new states will disrupt nucleosynthesis unless their mass is above $\sim 10$~MeV~\cite{PhysRevD.70.043526} and there are bounds from the number of relativistic degrees of freedom
in the early Universe~\cite{Boehm:2013jpa}. There also exist constraints on the scattering of $\chi$ with neutrinos from
the damping of structure in the Universe~\cite{Mangano:2006mp,Wilkinson:2014ksa}, though this only applies if $\chi$ forms the dominant component of dark matter (set by the size of $\sigma v$).

\begin{figure}[t]
\centering
\includegraphics[width=0.49\textwidth]{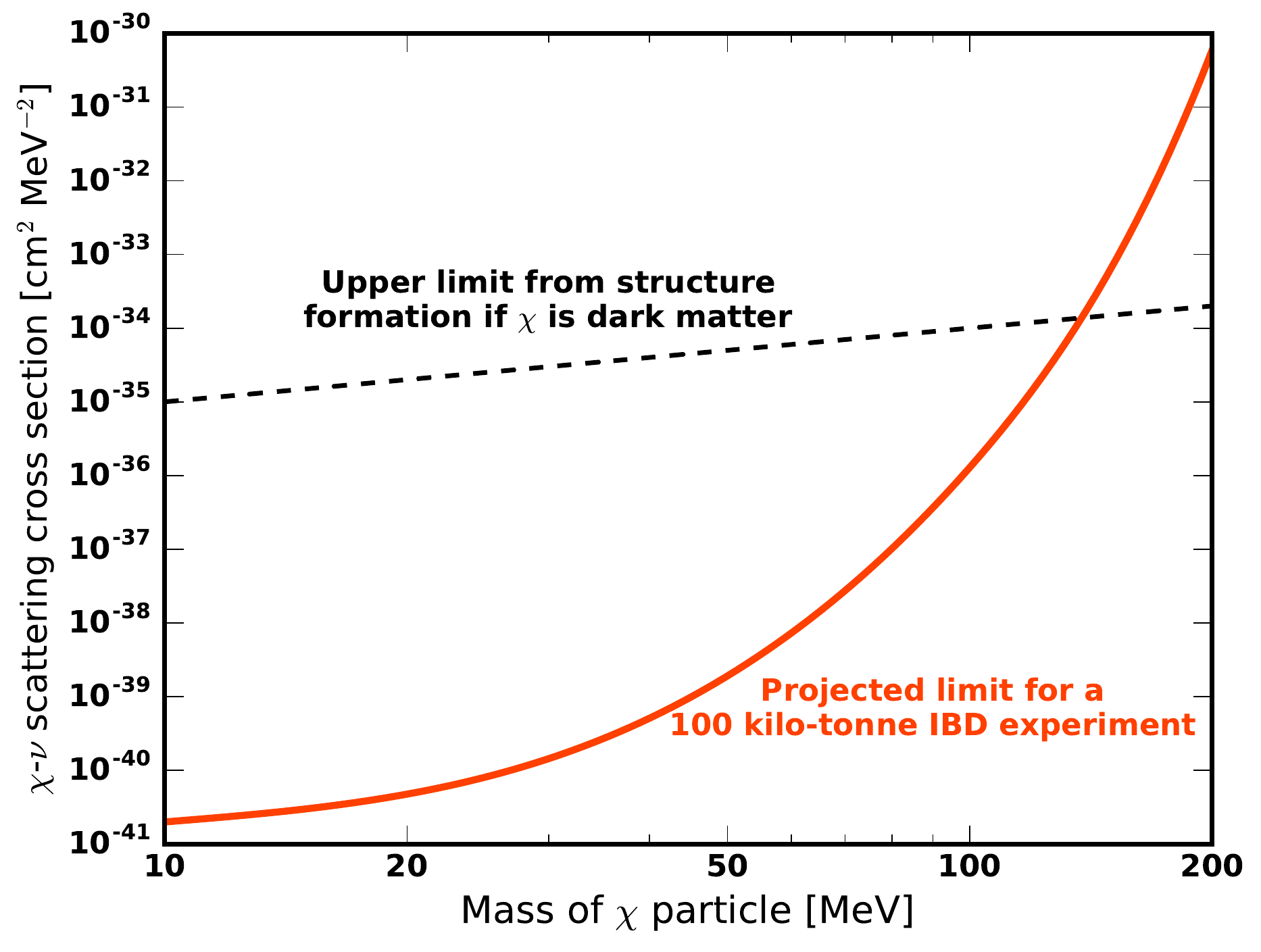}
\caption{Comparison of our projected upper limit on $\sigma_{\chi \nu}$ at $90\%$ confidence for a 100 kilo-tonne experiment looking for neutrinos through inverse beta decay (IBD) capture, with the upper limit from requiring that dark matter scattering
with neutrinos does not destroy small-scale structure in the Universe~\cite{Wilkinson:2014ksa,Mangano:2006mp}. The latter only applies if $\chi$ forms the dominant component of the dark matter.}
\label{fig:lims_mass}
\end{figure}

In figure~\ref{fig:lims_mass} we show our projected limit for an experiment with a mass of 100 kilo-tonnes detecting $\bar \nu_e$ via inverse beta decay for a range of $\chi$ masses above the bound from nucleosynthesis~\cite{PhysRevD.70.043526}.
Our projected limit is stronger than that from the damping of structure in the Universe~\cite{Mangano:2006mp,Wilkinson:2014ksa} for masses up to around 100~MeV, above which the density of $\chi$ in the hot star becomes too small
due to Boltzmann suppression to lead to any significant scattering with neutrinos, unless $\sigma_{\chi \nu}$ is very large.

\section{Conclusion \label{sec:conc}}
If a supernova event were to occur in our own galaxy then a large number of neutrino events should be detected on Earth in the space of a few seconds.  The majority of the neutrinos are produced thermally with a temperature depending on their flavour: electron neutrinos scatter the most with baryonic matter and so have the smallest temperature,
while electron anti-neutrinos are slightly hotter and the other heavy flavours hotter still~\cite{Sigl:1994da,Keil:2002in,Yuksel:2007mn}. 

In this work we have considered the prospect of additional MeV-mass `dark' particles $\chi$ being produced thermally during the supernova, which scatter with neutrinos~\cite{Sigl:1994da,Fayet:2006sa}. We have focused on the effect these new states would have on the
temperatures of the different neutrino flavours due to scattering, with cross section $\sigma_{\chi \nu}$. This leads to an extra term in the diffusion coefficient $D$ i.e. equation~(\ref{eqn:diff_chi}), which alters the hierarchy
of the temperatures for each flavour, as shown in figure~\ref{fig:temps_plot}. However quantifying this effect is made difficult by the uncertainty in the modelling of the density and temperature of the star during thermal production.

Our main emphasis has been on predicting the ability of current and future experiments to be able to measure the neutrino spectrum with
enough precision to place constraints on $\sigma_{\chi \nu}$. We show some example spectra in figure~\ref{fig:spectra_sigmas}. 
Since the temperatures of the different neutrino flavours depend sensitively on the model for the star during thermal production of neutrinos (and potentially $\chi$ also), we have performed a Bayesian analysis where we marginalise
over the parameters of a simple analytic stellar model (equations (\ref{eqn:nbprof}) and (\ref{eqn:Tprof})).
We have considered experiments looking at coherent nuclear scattering, such as direct detection experiments, and those looking for $\bar \nu_e$ via inverse beta decay
capture, such as Super and Hyper Kamiokande~\cite{Fukuda2003418,Abe:2011ts}.

As can be seen in figure~\ref{fig:lims}, we have found that Super Kamiokande could place an upper limit at the level of $\sigma_{\chi \nu} \sim 10^{-40} \cdot (T / \mathrm{MeV})^2$~cm$^2$ for 10~MeV mass $\chi$ particles and Hyper Kamiokande could be able to set a limit an order of magnitude stronger still.
Furthermore as shown in figure~\ref{fig:lims_mass} our projected limits are potentially the strongest constraint on $\chi$ scattering with neutrinos up to a mass around 100~MeV.
However these projections are essentially the `best case scenario' in the sense that we assume there are no additional effects on top of our simplified model for the star during neutrino production which could bring the temperatures
of $\bar \nu_e$ and $\bar \nu_{\mu / \tau}$ closer together~\cite{PhysRevLett.104.251101,Dasgupta:2008cd,Dasgupta:2009mg}. Such effects could weaken these limits considerably.

Indeed our analysis was performed using a simple empirical parametrisation to model the star during neutrino production, in order to understand how the uncertainties in this model impact our ability to constrain $\sigma_{\chi \nu}$. 
It is likely that there are further effects which will alter the expected temperatures of the neutrino species~\cite{PhysRevLett.104.251101,Keil:2002in,Buras:2005rp,Mirizzi:2015eza,Raffelt:2001kv,Dasgupta:2008cd,Dasgupta:2009mg}. However we expect that the value of $\sigma_{\chi \nu}$
where the temperatures of the different flavours become equal will remain fairly robust to additional modelling uncertainties. 

For detectors looking at coherent nuclear scattering the limits should in principle be more robust to such effects, since they will be able to measure all neutrino flavours through neutral-current scattering. However experiments
which exploit this channel, for example dark matter direct detection experiments~\cite{Baudis:2012bc,Baudis:2013qla,Billard:2013qya,Monroe:2007xp}, are generally much less massive than Super Kamiokande and so can not exploit as many statistics to measure the neutrino spectrum. This motivates
construction of an 100 tonne mass liquid-xenon experiment or similar as potentially the ultimate neutral-current supernova detector, with regards to effects from $\chi$-$\nu$ scattering.

\section*{Acknowledgements}
The author thanks John Beacom and Malcolm Fairbairn for helpful comments and suggestions.
The research leading to these results has received funding from the European Research Council through the project DARKHORIZONS under the European Union's Horizon 2020 program (ERC Grant Agreement no.648680).

\end{document}